\documentclass[twocolumn,amsmath,amssymb,aps,prb,floatfix,a4paper]{revtex4}
\usepackage[pdftex]{graphicx,color}
\usepackage{times}
\usepackage{subfigure}
\usepackage[latin1]{inputenc}
\usepackage[T1]{fontenc}
\newcommand{\V}{\boldsymbol}

\renewcommand\Re{\operatorname{\mathfrak{Re}}}
\renewcommand\Im{\operatorname{\mathfrak{Im}}}

\begin{document}

\title{Decoherence in quantum dots due to real and virtual transitions: a non-perturbative calculation\\}

\author{Thomas Grange}
\email[Electronic address:~]{thomas.grange@wsi.tum.de}

\affiliation{Laboratoire Pierre Aigrain, Ecole Normale Sup\'erieure, CNRS, 24 rue Lhomond, 75005 Paris, France} 
\affiliation{Walter Schottky Institut, Technische Universität München, Am Coulombwall 3, 85748 Garching, Germany}

\date{\today}

\begin{abstract}

We investigate theoretically acoustic phonon induced decoherence in quantum dots. We calculate the dephasing of fundamental (interband or intraband) optical transitions due to real and virtual transitions with higher energy levels.
Up to two acoustic phonon processes (absorption and/or emission) are taken into account simultaneously in a non-perturbative manner.
An analytic expression of acoustic phonon induced broadening is given as a function of the electron-phonon matrix elements and is physically interpreted. The theory is applied to the dephasing of intersublevel transitions in self-assembled quantum dots.

\end{abstract}

\maketitle

Understanding decoherence in semiconductor quantum dots (QDs) is of prime importance in the context of their potential application in quantum information.
Dephasing of the interband transitions in semiconductor QDs have been extensively studied in the past few years both experimentally and theoretically.
It has been show that the interaction with lattice vibrations (phonons) is a major source of dephasing \cite{borri01,takagahara99,krummheuer02} of the optical polarization. 
From the theoretical point of view, interaction between the electronic levels and acoustic phonon can be decomposed into diagonal and off-diagonal terms. The diagonal part gives rise to acoustic phonon sidebands \cite{besombes01,favero03}. This sidebands are responsible of a rapid and partial decay of the polarization which is now well understood \cite{borri01,krummheuer02,vagov04}.
Different mechanisms have been put forward in order to explain the broadening of the zero phonon line (ZPL) \cite{muljarov04,borri05,machnikowski06,muljarov07}. In particular, Muljarov and Zimmermann \cite{muljarov04} have proposed a mechanism of virtual transitions triggered by the off-diagonal acoustic phonon interaction.

More recently, dephasing of the $s$-$p$ intersublevel transition has been studied in QDs \cite{zibik08}. While at low temperature (10K) dephasing has been shown to be limited by anharmonic polaron decay mechanism \cite{polaronrelax}, at higher temperature both real and virtual transitions between the two $p$ states were identified as responsible for the broadening of the ZPL.

Virtual transitions refer to two-phonon processes of simultaneous absorption and emission of phonons with same energy but different momentum (\textit{i.e.} elastic diffusion of phonons) due to non-resonant phonon couplings (off-diagonal electron-phonon interaction) with an higher excited state.
This process contributes to the decay of the coherence in addition to other dephasing processes such as real transitions.

In self-assembled quantum dot, phonon absorption or emission is efficient only if the energy separation between levels is of the order of a few meV, corresponding to acoustic phonon wavelength of the order of the dot size. However, energy separation between electronic levels is generally larger so that real transitions (single phonon absorption or emission) are in general not efficient. This effect is known as the acoustic phonon bottleneck effect \cite{bockelmann90}. Virtual transitions instead (because of their non-resonant character) are less sensitive to the energy detuning between electronic levels and provide a contribution to the dephasing in a larger range of electronic energy separation.

To date, two-phonon processes in QDs have only be treated perturbatively with respect to the electron-phonon matrix elements. Takagahara has used a self-energy expansion \cite{takagahara99}, while Muljarov and co-workers have used a cumulant expansion \cite{muljarov04,muljarov05}.
In the present article, we present a calculation of the dephasing in quantum dots taking into account up to two-phonon processes in a non-pertubative manner.
Contrary to previous approach, our calculation includes a resummation of all the diagrams that involve states which differs from the initial reservoir by 1 or 2 phonons.
This is achieved using projection operators \cite{mower66,cohen,mukamel} twice consecutively. 
An analytic formula is obtained as a function of the electron-phonon matrix elements, which is physically interpreted and allows to identify the contribution of the different phonons to the dephasing processes as a function of their energy.

\section{Theory of two-phonon processes induced dephasing}

We consider a three level system:
the ground state is denoted $|g\rangle$, while $|e_0\rangle$ and $|e_1\rangle$ are the first and second excited states. We study here the coherence of the transition between $|g\rangle$ and $|e_0\rangle$. 
We consider the following Hamiltonian:
\begin{equation}
H= H_{\text{e}} + H_{\text{ph}}+ V_{\text{e-ph}}
\end{equation}
where $H_{\text{e}}$ is the electronic part, $H_{\text{ph}}$ is the acoustic phonon Hamiltonian, and $V_{\text{e-ph}}$ represents the electron-phonon interaction.

\begin{subequations}
\begin{equation}
H_{\text{e}}= \varepsilon_g |g\rangle \langle g | + \varepsilon_0 |e_0\rangle \langle e_0 | + (\varepsilon_0+\Delta) |e_1\rangle \langle e_1 |
\end{equation}
\begin{equation}
H_{\text{ph}}= \sum_i \varepsilon_i a^+_{\V{q}} a_{\V{q}}
\end{equation}
\begin{equation}
V_{\text{e-ph}}=\sum_{n,m}\sum_{\V{q}}M_{\V{q}}^{nm}(a_{\V{q}}+a_{\V{-q}}^{+})\vert n\rangle\langle m\vert\label{coupling}
\end{equation}
\end{subequations}
where the sums over $n$ and $ m$ run over $g$, $e_0$ and $e_1$. 
In quantum dots, the transition energy $\varepsilon_0-\varepsilon_g$ is usually large compared to the energy separation $\Delta$ between the first two excited states $|e_0\rangle$ and $|e_1\rangle$ (in both intraband and interband cases). As a consequence, the off-diagonal acoustic phonon couplings affecting the ground state ($M_{\V{q}}^{ge_0}$ and $M_{\V{q}}^{ge_1}$) can be neglected.
In addition, diagonal interaction in the $g$ state (\textit{e. g.} for intraband transitions) is removed through a Huangh-Rhys transformation of the form $a_q \rightarrow a_q - M_q^{gg*}/\varepsilon_q $. The new phonon modes are thus the eigenmodes of the crystal vibrations when the system is in the ground state $|g\rangle$.
In the following, the total Hamiltonian will be split as $H=H_0+V_{\text{e-ph}}$ where the non-interacting part is $H_0=H_{\text{e}}+H_{\text{ph}}$.

We are interested in the decay of the optical polarization of the $g$-$e_0$ transition due to the interaction of the phonon reservoir assumed to be at thermal equilibrium.
We consider an initial eigenstate of the form $|g,\phi_0\rangle$, where $|\phi_0\rangle$ denotes 
an initial phonon state
$|\phi_0 \rangle =  |n_{1}, n_{2},...,n_{N_{cr}}\rangle$ where
$n_{i}$
is the occupancy of the LA-phonon mode $i$ ($n_{i}=\langle \phi_0 | a_i^+a_i|\phi_0 \rangle$) and $N_{cr}$ is the number of atom in the crystal. The limit of an infinite crystal is considered ($N_{cr}\rightarrow +\infty$). In this limit the integer values of the $n_{i}$ are distributed according to the Bose-Einstein function $N(\varepsilon)=1/(e^{\varepsilon/k_{B}T}-1)$.
The dipole operator of the $g$-$e_0$  transition is defined as $d=|g\rangle \langle e_0| + |e_0\rangle \langle g|$.
The linear polarization  (the linear component of the polarization in response of a delta pulse excitation) is given by \cite{mukamel}:
\begin{equation}
P(t) = i \langle g,\phi_0 | d(t) d |g,\phi_0\rangle \rangle
\label{polarization}
\end{equation}
where $ d(t)=e^{iHt/\hbar} d e^{-iHt/\hbar}$.

As $|e_0,\phi_0\rangle$ is an eigenstate of the full Hamiltonian $H$, Eq.~\ref{polarization} reduces to:
\begin{equation}
P(t)= i e^{i\varepsilon_g t/\hbar} \langle e_0,\phi_0 | e^{-iHt/\hbar} |e_0,\phi_0\rangle
\end{equation}

It is known that diagonal acoustic phonon interaction in QDs leads to phonon sidebands that are responsible for a partial polarisation decay.
A Huangh-Rhys transformation is made with new phonon modes defined by  $b_q = a_q + M_q^{e_0e_0*}/\varepsilon_q$. The new occupancy states are then denoted $|\widetilde{n_i}\rangle=(b_i^+)^{n_ i}/\sqrt{n!}|0\rangle$.
As shown in Appendix, 
$P(t)$ is given by:
\begin{equation}
P(t)=ie^{-i\omega t} g(t) P_Z(t)
\end{equation}
\begin{equation}
P_Z(t)=\langle e_0,\widetilde{\phi_0} | e^{-iHt/\hbar}|e_0,\widetilde{\phi_0}\rangle
\end{equation}
where $g(t)$ is defined in the appendix and corresponds to the polarization decay due to the sidebands only, $|\widetilde{\phi_0}\rangle = |\widetilde{n_1},\widetilde{n_2}, ...,\widetilde{n_{N_{cr}}} \rangle$ ($|\widetilde{\phi_0}\rangle$ has the same phonon occupancies as $|\phi_0\rangle$ and is involved in the ZPL component of the transition),
and $\omega$ is the frequency of the $g$-$e_0$ ZPL optical transition. The energy origin is from now taken at the $|e_0,\widetilde{\phi_0}\rangle$ level.

The remaining task is the calculation of $P_Z(t)$.
The diagonal electron-phonon matrix elements after the Huangh-Rhys transformtion reads
$\widetilde{M^{e_0e_0}}=0$ and $\widetilde{M^{e_1e_1}}=M^{e_1e_1}-M^{e_0e_0}$.
As the calculation of $P_Z(t)$ will be made in the new phonon mode basis, $|\widetilde{\phi_0}\rangle$ will be denoted $|\phi_0\rangle$ in the following in order to keep the notations simple.
For $t>0$, the evolution operator can be expressed as:
\begin{equation}
e^{-iHt/\hbar}  = -\frac{1}{2 i \pi}\int_{-\infty}^{+\infty}dE e^{-iE\tau/\hbar} G(E)
\end{equation}
where $G(E)=1/(E^+-H)$ denotes the retarded Green function ($E^+=E+i\eta$ with $\eta \rightarrow 0^+$).
In order to evaluate this term, we use the method of projection operator \cite{mower66,cohen,mukamel}.
Let us recall a property of this method. If $P$ is a projection operator on a subspace of interest and $Q=1-P$, the restriction of the Green function $G$ to the subspace defined by $P$ reads:
\begin{equation}
PG(E)P = \frac{P}{E^+ - PH_0P - PR(E)P}
\label{thp}
\end{equation}
where the self energy operator $R$ is given by:
\begin{equation}
R(E) = V + V\frac{Q}{E^+ - QH_0Q - QVQ}V \\
\label{opdeplacement}
\end{equation}

This property will be used twice in the following.
First, $P$ is defined as the projection operator on the $|\phi_0\rangle$ phonon state, while $Q$ is defined as $Q=1-P$.
As the energy origin is taken at the $|e_0,\phi_0\rangle$ level, we have:

\begin{equation}
\langle e_0,\phi_0 | G(E)|e_0,\phi_0\rangle =  \frac{1}{E-D(E)+iI(E)}
\label{greengeneral}
\end{equation}
where $I(E)=\Im\langle e_0,\phi_0 | R(E)|e_0,\phi_0\rangle$ ($\Im$ stands for imaginary part) and $D(E)=\frac{1}{2\pi} \mathcal{P} \int \text{d}E' I(E')/(E-E') $.
The energy displacement $D(E)$ is usually found to be small for acoustic phonon-electron interaction in QDs and can be neglected compared to the energy separation between the QD electronic levels that are considered, so that $D(E) = 0$ will be taken in the following. Also note that the non-diagonal coupling $\langle e_0\phi_0|R|e_1\phi_0\rangle$ is found to vanish in the following.
In addition, we will have often  $\text{d}I/\text{d}E \ll 1$ so that $I(E)$ can be replaced by $I(0)=\Gamma_{0}$ in Eq.~\ref{greengeneral}, \textit{i.e.} the ZPL lineshape can be approximated by a Lorentzian. Hence the polarization corresponding to the ZPL will exhibit a mono-exponential decay of the form $P_{Z}(t)= e^{-\Gamma_{0}t/\hbar}$.

\subsection{One-phonon processes.}

If we consider only one phonon processes, the term $QVQ$ cancels in Eq.~\ref{opdeplacement}, so that

\begin{equation}
\Gamma_{0} =\gamma_{od}(-\Delta)
\end{equation}

\begin{equation}
\gamma_{od}(E) =
\begin{cases} (N(E)+1)\Gamma_{od}(E) & \text{if $E>0$} \\ N(-E)\Gamma_{od}(-E) & \text{if $E<0$} \end{cases}
\end{equation}

\begin{equation}
\Gamma_{od}(E) = \pi \sum_{\phi_i} |M_{i}^{e_0e_1}|^2 \delta(E - \varepsilon_i)
\label{gammaod}
\end{equation}

where $N(\varepsilon)=1/(e^{\varepsilon/k_{B}T}-1)$ is the Bose
occupation number.
The decay rate $\Gamma_{0}/\hbar$ of the ZPL polarization corresponds here to the Fermi golden rule for acoustic phonon absorption ($\Gamma_{0}=N(\Delta)\Gamma_{od}(\Delta)$ if $\Delta>0$) or phonon emission ($\Gamma_{0}=(N(-\Delta)+1)\Gamma_{od}(-\Delta)$ if $\Delta<0$).
The population decays with the rate  $2\Gamma_{0}/\hbar$.

\subsection{Two-phonon processes.}

In order to take into account both real and virtual transitions, we have to treat simultaneously one and two phonon processes.
In the calculation of $P_Z(t)=\langle e_0,\phi_0 | e^{-iHt/\hbar}|e_0,\phi_0\rangle$, we will restrict the Hamiltonian to phonon states that differ by 2 phonons or less from $|\phi_0\rangle$. Within this framework, the calculation will be made non-perturbatively. This is achieved in the following using different projectors. 
In order to develop Eq.~\ref{opdeplacement}, we define a new retarded Green function 
\begin{equation}
G'(E)=\frac{Q}{E^+ - QH_0Q - QVQ}
\end{equation}
Introducing two completeness relation in Eq.~\ref{opdeplacement} gives:
\begin{equation}
\langle e_0,\phi_0 | R(E) |e_0,\phi_0\rangle = \sum_{\phi_a}\sum_{\phi_b} V_{0a}V_{b0} G'_{ab}(E)
\label{decomp}
\end{equation}

where $V_{\alpha\beta}= \langle e_0,\phi_{\alpha} | V |e_1,\phi_{\beta}\rangle $ and $G'_{\alpha\beta}(z) = \langle e_1,\phi_{\alpha} | G'(z) | e_1,\phi_{\beta} \rangle$.
The term $V_{0a}$ ($V_{b0}$) is non-zero only if $\phi_a$ ($\phi_b$) differs from $\phi_0$ by only one phonon (absorption or emission). Hence each $\phi_a$ ($\phi_b$) contributing to the summation is either of the form $|+i\rangle\equiv a_i^+|\phi_0\rangle/\sqrt{\langle \phi_0 | a_i a_i^+ | \phi_0 \rangle}$ or $|-i \rangle \equiv a_i|\phi_0\rangle/\sqrt{\langle \phi_0 | a_i^+ a_i | \phi_0 \rangle}$.
This summation involves diagonal as well as off-diagonal terms for $G'$, which can be expressed separately:

\begin{equation}
\langle e_0,\phi_0 | R(E) |e_0,\phi_0\rangle = \sum_{\phi_a} |V_{0a}|^2 G'_{aa} +  \sum_{\phi_a}\sum_{\phi_b\neq\phi_a} V_{a}V_{b}^* G'_{ab}
\label{eqgprime}
\end{equation}

We have now to calculate the matrix elements (diagonal and off-diagonal) of the operator $G'=1/(E^+ - QH_0Q - QVQ)$. This operator is similar to the original Green operator except that $H_0$ and $V$ have been replaced by $QH_0Q$ and $QVQ$.
Similarly, we define new projection operators: $P'$ is the projection operator on the subspace defined by by phonon states differing from $|\phi_0\rangle$ by one phonon  
($\{\phi_i$ such as $V_{0i} \neq 0 \}$) and $Q'=1-P'$.
Applying  Eqs.~\ref{thp} and \ref{opdeplacement} once again leads to:

\begin{equation}
P'G'(E)P' =  \frac{P'}{E^+ - P'H_0P' - P'R'(E)P'}
\label{res2}
\end{equation}

\begin{equation}
P'R'(E)P' = P' V G''(E) V P'
\label{rprime}
\end{equation}
where $G''(E)=Q''/(E^+ - Q''H_0Q'' - Q''VQ'')$ with $Q''=QQ'=1-P-P'$.
$Q''$ is the projector on states that differ by at least 2 phonon from the initial state $|\phi_0\rangle$. As the calculation is made up to 2 phonon processes, $Q''=P''$ where $P''$ is the projector on states differing by exactly 2 phonons from $|\phi_0\rangle$ (states differing by 3 phonons or more from $|\phi_0\rangle$ are not taken into account). As a consequence the term $Q''VQ''$ in $G''$ vanishes since $V$ couples states that differ by one phonon.

The diagonal matrix elements of $R'$ read:
\begin{equation}
R'_{aa} = \sum_{|f\rangle} |V_{af}|^2 G''_f
\end{equation}
where $V_{af} = \langle e_1,\phi_a | V | f \rangle  $ 
and $G''_f= \langle f | G'' | f \rangle  $. The $| f \rangle $ states differs necessarly by 2 phonons from $|\phi_0\rangle$.
The non-diagonal matrix elements of $R'$ read:
\begin{equation}
R'_{ab} = \sum_{|f\rangle} V_{af} V_{fb} G''_f
\end{equation}

In order $V_{af} V_{fb}$ to be non-zero in this equation, the phonon state of $|f\rangle$ should be of the form $| \pm i , \pm j \rangle $ given that $\phi_a$ and $\phi_b$ are of the form $|\pm i \rangle$ and $|\pm j \rangle$
($|f\rangle=|e_{0/1},\phi_0\rangle$ is forbidden by the $Q''$ projector).

The non-diagonal matrix elements of $R'$ are of the order of $\mathcal{O}(1/N_{cr})$ since they involve a product of two terms in $V$, while the diagonal matrix elements of $R'$ involve a summation over the $N_{cr}$ phonon modes and are consequently of the order of $\mathcal{O}(1)$.
In the following, we decompose $P'R'P'$ into diagonal $R'_{d}$ and non-diagonal $R'_{od}$ contributions.
In the matrix elements of $G'$, we need to keep only terms up to the first order in $R'_{od}$ (\textit{i.e.} in $\mathcal{O}(1/N_{cr})$) since the higher order terms give a vanishing contribution in Eq.~\ref{eqgprime}.
Hence inverting equation~\ref{res2} gives:
\begin{equation}
P'G'P'= G'_d + G'_d R'_{od} G'_d
\end{equation}
where $G'_d= P'/(z-P'H_0P'-R'_{d})$ is the diagonal part of P'G'P'.
The non-diagonal elements of $G'$ read:
\begin{equation}
G'_{ab}= G'_{aa}R'_{ab}G'_{bb}
\end{equation}
where $R'_{\alpha\beta} = \langle e_1,\phi_{\alpha} | R' | e_1,\phi_{\beta} \rangle$
Following the assumption that energy displacements are negligible, we keep only the imaginary part of $G''_f(E)\simeq-i\pi \delta(E-E_f)$.
We keep also only the imaginary part in the matrix elements of $R'$.
The imaginary part of $G'_{aa}$ can then be expressed as:
\begin{equation}
\Im ~ ( G'_{aa} )= - R'_{aa} |G'_{aa}|^2 = - |G'_{aa}|^2 \sum_{f} |V_{af}|^2 G''_f
\end{equation}

Now we can express the imaginary part of the self-energy:

\begin{widetext}

\begin{equation}
-i \Gamma_0
= - \sum_{\phi_a} |V_{0a}|^2 |G'_{aa}|^2 \sum_{f} |V_{af}|^2 G''_f  - \sum_{\phi_a}\sum_{\phi_{b\neq a}}\sum_{f} \Re ~ (V_{0a}G'_{aa}V_{af}V_{fb}G'_{bb}V_{b0}) G''_{f}
\end{equation}

\end{widetext}

where $\Re$ stands for real part.
The phonon part of $|f\rangle$ differs from the initial phonon state $|\phi_0\rangle$ by 2 phonon occupancies. $|f\rangle$  being of the form $|e_{0/1},\pm i, \pm j \rangle$ (contributions of the form $|e_{0/1},\pm 2 i \rangle$ gives vanishing contribution in the limit $N_{cr}\rightarrow \infty$), one can always define the two state $|\phi_a\rangle=|\pm i\rangle$ and $|\phi_b\rangle=|\pm j\rangle$ so that $|e_1,\phi_a\rangle$ and $|e_1,\phi_b\rangle$ are the only states coupled to $|e_0,\phi_0\rangle$ and $|f\rangle$ simultaneously.
As a consequence, the double summation over the phonon modes $|\phi_a\rangle$ and $|\phi_b\rangle$
can be replaced by a simple summation over the $|f\rangle$ states:

\begin{equation}
\Gamma_0
= \pi \sum_{f}  \left\vert V_{0a}G'_{aa}V_{af}  +  V_{0b} G'^{*}_{bb} V_{bf} \right\vert^2 \delta(\varepsilon_f)
\end{equation}

The above formula can be expressed as a sum of two contribution,which corresponds respectively to the electronic part of $|f\rangle$ being either $e_0$ or $e_1$:
\begin{equation}
\Gamma_0=\Gamma_0^{(\alpha)}+\Gamma_0^{(\beta)}
\end{equation}

We define $V_{a}^{od}=\langle e_0, \phi_0 | V| e_1, \phi_a \rangle$ and  $V_{a}^{d}=\langle e_1, \phi_0 | V| e_1, \phi_a\rangle$.
If  $|f\rangle$  is of the form $|e_{0},\pm i, \pm j \rangle$, we have $V_{0a}=V_{a}^{od}$, $V_{af}=V_{b}^{nd*}$, $V_{0b}=V_{b}^{od}$ and $V_{bf}=V_{a}^{nd*}$,  so that $\Gamma_0^{(\alpha)}$ reads:

\begin{equation}
\Gamma_0^{(\alpha)}= \frac{\pi}{2} \sum_{\phi_a}\sum_{\phi_b}  |V_{a}^{od}|^2|V_{b}^{od}|^2 |G'_{aa} +  G'^{*}_{bb}|^2 \delta(\varepsilon_a+\varepsilon_b)
\label{gammaalpha}
\end{equation}

The $\Gamma_0^{(\beta)}$ term corresponding to $|f\rangle$ in the $e_1$ electronic state reads:

\begin{equation}
\Gamma_0^{(\beta)}
= \frac{\pi}{2} \sum_{\phi_a}\sum_{\phi_b}  \left\vert V_{a}^{od}V_{b}^{d} G'_{aa} +  V_{b}^{od}  V_{a}^{d} G'^{*}_{bb}\right\vert^2 \delta(\Delta + \varepsilon_a + \varepsilon_b)
\label{gammabeta}
\end{equation}

\begin{figure}
\begin{centering}
\includegraphics[width=0.45\textwidth]{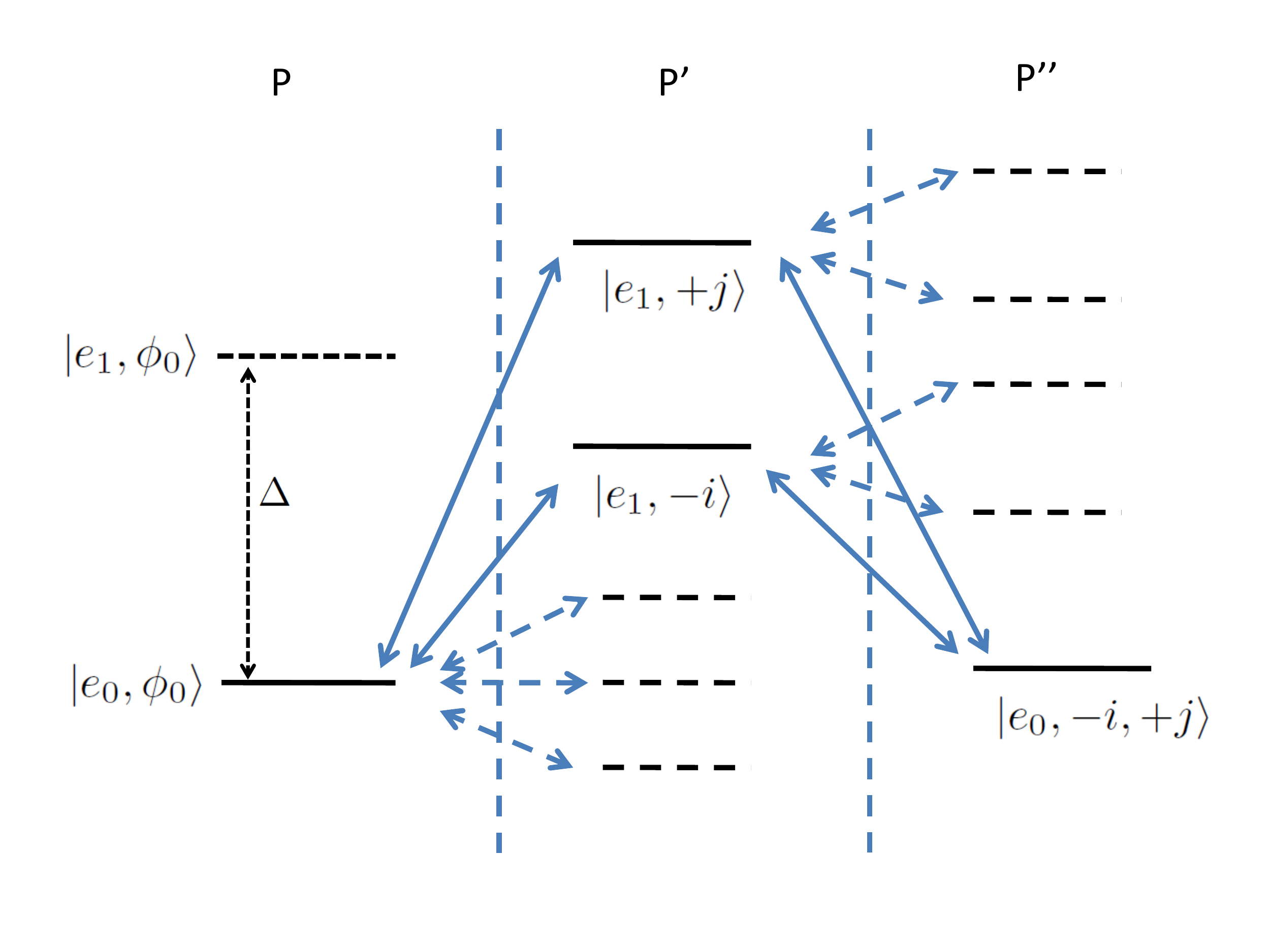}
\par\end{centering}

\caption{(Color online) Schematic of the electron-phonon levels involved in the dephasing processes. Levels are classified in three columns according to the number of phonon occupancies they differ from $\phi_0$: the projector operator $P$, $P'$, and $P''$ corresponds respectively to zero, one and two-phonon difference from the initial phonon state.
The levels in solid lines illustrate coherent couplings involved in two-phonon virtual processes.}
\label{schematic}
\end{figure}

To evaluate these terms we need first to express the diagonal matrix elements of $R'$ which are involved in the diagonal matrix elements of $G'$. They can be expressed as the sum of two contributions:
\begin{equation}
R'_{aa}(i0^+) = -i \gamma_{od}(-\varepsilon_a)
-i \gamma_{d}(-\Delta-\varepsilon_a)
\end{equation}
where the following quantity are defined:
\begin{equation}
\gamma_{od}(E) = \pi \sum_{\phi_a} |V_{a}^{od}|^2 \delta(E - \varepsilon_a)
\end{equation}

\begin{equation}
\gamma_{d}(E) = \pi \sum_{\phi_a} |V_{a}^{d}|^2 \delta(E - \varepsilon_a)
\end{equation}

$\gamma_{od}$ and $\gamma_{d}$ corresponds respectively to the broadening due to non-diagonal and diagonal interaction within the $e_1$ state. Using the Bose-Einstein distribution function $N(E)=1/\left(e^{E/kT}-1\right)$, they can be expressed as
\begin{equation}
\gamma_{od/d}(E) =
\begin{cases} (N(E)+1)\Gamma_{od/d}(E) & \text{if $E>0$} \\ N(-E)\Gamma_{od/d}(-E) & \text{if $E<0$} \end{cases}
\end{equation}

where $\Gamma_{od}(E)$ was defined in Eq.~\ref{gammaod} and 

\begin{equation}
\Gamma_{d}(E) = \pi \sum_{\phi_i} |M_{i}^{e_1e_1}-M_{i}^{e_0e_0}|^2 \delta(E - \varepsilon_i)
\end{equation}

The diagonal matrix elements of $G'$ involved in Eqs.~\ref{gammaalpha} and \ref{gammabeta} can now be expressed as:
\begin{equation}
G'_{aa}(0) = \frac{1}{-\Delta-\varepsilon_a+i \gamma_{od}(-\varepsilon_a)
+i \gamma_{d}(-\Delta-\varepsilon_a)}
\end{equation}

In the following, we will consider first only off-diagonal interactions, and then both diagonal and off-diagonal electron-phonon interactions.

\section{Off-diagonal interactions}

In this section, we consider only off-diagonal interaction in $V_{\text{e-ph}}$.
In this case, only the term arising from $\Gamma_0^{(\alpha)}$ remains:

\begin{equation}
\Gamma_0
= \frac{\pi}{2} \sum_{a}\sum_{b}  \delta(\varepsilon_a+\varepsilon_b) |V_{a}^{od}|^2|V_{b}^{od}|^2 |G'_{a} +  G'^{*}_{b}|^2
\end{equation}

Using $\varepsilon_b=-\varepsilon_a$, we obtain

\begin{widetext}

\begin{equation}
\Gamma_0
=\frac{\pi}{2} \sum_{a}\sum_{b}  \delta(\varepsilon_a+\varepsilon_b) |V_{a}^{od}|^2|V_{b}^{od}|^2 
\frac{4\Delta^2+(\gamma_{od}(-\varepsilon_a)-\gamma_{od}(\varepsilon_a))^2}
{((\Delta+\varepsilon_a)^2+(\gamma_{od}(-\varepsilon_a))^2)((\Delta-\varepsilon_a)^2+(\gamma_{od}(\varepsilon_a))^2)}
\end{equation}

In the continuum limit for phonons ($N_{cr}\rightarrow \infty$), inserting an integral over the energy leads to: 

\begin{equation}
\Gamma_0
= \frac{1}{2\pi} \int_{-\infty}^{+\infty} d\varepsilon \gamma_{od}(\varepsilon)\gamma_{od}(-\varepsilon)
\frac{4\Delta^2+(\gamma_{od}(-\varepsilon)-\gamma_{od}(\varepsilon))^2}
{((\Delta+\varepsilon)^2+\gamma_{od}^2(-\varepsilon))((\Delta-\varepsilon)^2+\gamma_{od}^2(\varepsilon))}
\end{equation}

In order to explicit the temperature dependence, $\gamma_{od}$ can be replaced by its expression as function of $\Gamma_{od}$ and the Bose distribution function $N(\varepsilon)$.

\begin{equation}
\Gamma_0
= \frac{1}{\pi} \int_{0}^{+\infty} d\varepsilon N(\varepsilon)(N(\varepsilon)+1)\Gamma_{od}^2(\varepsilon)
\frac{4\Delta^2+\Gamma_{od}^2(\varepsilon)/4}
{\left[(\Delta+\varepsilon)^2+\left[N(\varepsilon)\Gamma_{od}(\varepsilon)\right]^2\right]\left[(\Delta-\varepsilon)^2+\left[(N(\varepsilon)+1)\Gamma_{od}(\varepsilon)\right]^2\right]}
\end{equation}

As $\Delta$ is usually large compared to $\Gamma_{od}(\varepsilon)$, a simplified expression can often be used (\textit{e.g.} in ref.~\onlinecite{zibik08}):
\begin{equation}
\Gamma_0
= \frac{1}{\pi} \int_{0}^{+\infty} d\varepsilon \frac{4\Delta^2}{(\Delta+\varepsilon)^2} 
\frac{N(\varepsilon)(N(\varepsilon)+1)\Gamma_{od}^2(\varepsilon)}
{(\Delta-\varepsilon)^2+\left[(N(\varepsilon)+1)\Gamma_{od}(\varepsilon)\right]^2}
\label{gsimple}
\end{equation}

Physically, this expression contains both the effect of real transitions due to one phonon processes and virtual processes involving two phonons.
 As discussed below, the peak around $\varepsilon=\Delta$ will be attributed to one-phonon absorption process, while the contribution of non-resonant phonon ($\varepsilon$ away from  $\Delta$) will be interpreted as virtual transitions, \textit{i.e.} an elastic diffusion of phonons without change in the electronic state.
The term $4\Delta^2/(\Delta+\varepsilon)^2$ appearing in the above equation can be understood as an interference term between two different path contributing to virtual processes: the intermediate state can be obtained either from phonon absorption or emission (see Fig.~\ref{schematic}).
These two path interfere constructively (destructively) for phonon energies $\varepsilon$ smaller (larger) than $\Delta$.

\section{Diagonal and off-diagonal interactions}

Including the diagonal interactions, the $\Gamma_0^{(\alpha)}$ term is given by:

\begin{equation}
\Gamma_0^{(\alpha)}= \frac{1}{2\pi} \int_{-\infty}^{+\infty} d\varepsilon \gamma_{od}(\varepsilon)\gamma_{od}(-\varepsilon)
\frac{4\Delta^2+(\gamma_{m}(\varepsilon)-\gamma_{m}(-\varepsilon))^2}
{\left[(\Delta+\varepsilon)^2+\gamma_{m}^2(-\varepsilon)\right]\left[(\Delta-\varepsilon)^2+\gamma_{m}^2(\varepsilon)\right]}
\label{galpha}
\end{equation}
where $\gamma_{m}(\varepsilon)=\gamma_{od}(\varepsilon)+\gamma_{d}(-\Delta+\varepsilon)$.
The term $\gamma_{m}(\varepsilon)$ plays a significant role only for $\varepsilon \simeq \pm \Delta$, which involves $\gamma_{d}$ evaluated around $0$ where it vanishes. As a consequence $\Gamma_0^{(\alpha)}$ is almost not affected by the inclusion of the diagonal interactions.

The term $\Gamma_0^{(\beta)}$ (which was vanishing in the absence of diagonal interactions) corresponds to real transitions assisted by 2-phonon absorption/emission, and depends on the quantity:

\begin{equation}
\left\vert V_{a}^{od}V_{b}^{d} G'_{a} +  V_{b}^{od}  V_{a}^{d} G'^{*}_{b}\right\vert^2 =
2 \sum_a \sum_b |V_{a}^{od}|^2|V_{b}^{d}|^2 |G'_{a}|^2
+2\sum_a \sum_b \Re( V_{a}^{od}V_{b}^{d} G'_{a} V_{b}^{nd*}  V_{a}^{d*} G'_{b})
\end{equation}

In many cases, such as for fundamental intraband or interband transitions in QDs, the second term vanishes because of the symmetry of the wavefunctions $\sum_q M_{q}^{e_1e_1} M_{q}^{e_0e_1*}=0$.
In such case, $\Gamma_0^{(\beta)}$  reads:

\begin{equation}
\Gamma_0^{(\beta)}= \pi \sum_{a}\sum_{b} \delta(\Delta + \varepsilon_a + \varepsilon_b) |V_{a}^{od}|^2|V_{b}^{d}|^2 
\frac{1}{(\Delta+\varepsilon_a)^2+\gamma_{m}^2(-\varepsilon_a)}
\label{gbeta}
\end{equation}

\end{widetext}

Inserting integrals in the above expression leads to:

\begin{equation}
\Gamma_0^{(\beta)} = \frac{1}{\pi} \int_{-\infty}^{+\infty} d\varepsilon 
\frac{\gamma_{od}(\varepsilon)\gamma_{d}(-\Delta-\varepsilon)}
{(\Delta+\varepsilon)^2+\gamma_{m}^2(-\varepsilon_a)}
\end{equation}

Note that this formula can be written as $f_c(-\Delta)$, where $f_c (E)= \int \text{d}\varepsilon \gamma_{od}(\varepsilon) A_s (E-\varepsilon)$ is the convolution between $\gamma_{od}$ (Fermi golden rule for one-phonon absorption/emission) and the function $A_s(\varepsilon)\simeq\gamma_{d}(\varepsilon)/\varepsilon^2$ describing the one-phonon phonon contribution to the sidebands spectrum of the $e_0$-$e_1$ transition.
In analogy with acoustic phonon sidebands of optical transition, the above formula can be interpreted as the apparition of phonon sidebands in a acoustic phonon assisted transition. As our calculation is made up to two-phonon processes, the sidebands taken into account involve only one-phonon process.
For optical transition, however, the apparition of sidebands does not change the radiative lifetime since the relative variation of photon energy is negligible on the energy scale of acoustic phonons. On the contrary, acoustic phonon sidebands modify the efficiency of acoustic phonon induced transitions through an additional contribution of multiphonon processes.

\section{Application: dephasing of intersublevel transition}

In order to illustrate the theory presented above, we study the dephasing of intersublevel transition in InAs/GaAs self-assembled quantum dots between the ground $s$-like state and the $p$-like first excited state (denoted $p_x$ here). The higher energy $p$-state is denoted $p_y$.
The contribution of acoustic-phonon to the dephasing is calculated according to the Eqs.~\ref{galpha} and \ref{gbeta}.
In Ref.~\onlinecite{zibik08}, the temperature dependence of the linewidth was presented, taking into account 2-phonon processes induced by off-diagonal interactions (Eq.~\ref{gsimple}). Here we take into account additionally the contribution of 2-phonon real transitions due to diagonal terms (Eq.~\ref{gbeta}). The QD parameters are the same as in Ref.~\onlinecite{zibik08}, with a $s$-$p_x$ transition energy of 53 meV.

\begin{figure}
\begin{centering}
\includegraphics[width=0.45\textwidth]{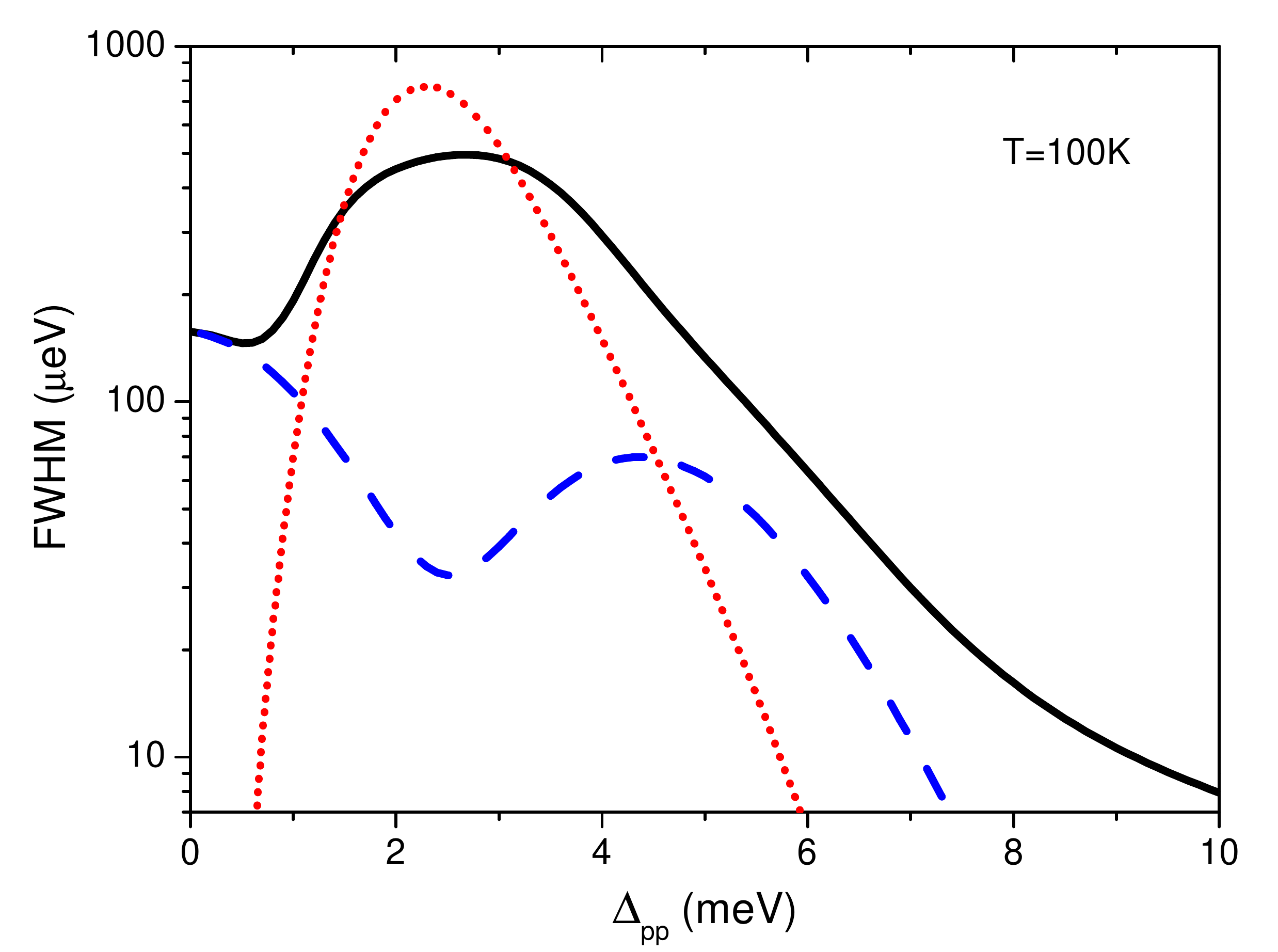}
\par\end{centering}

\caption{(Color online) Full width at half maximum (FWHM) $2\Gamma_0$ (full line) of the ZPL component of the $s$-$p_x$ intersublevel transition as a function of the energy detuning $\Delta_{pp}$ between the $p$-levels for T=100K. The Fermi golden rule for phonon absorption $2\gamma_{od}(-\Delta)$ is shown in dotted line. The contribution of real 2-phonon transitions $2\Gamma_0^{\beta}$ appears in dashed line.}
\label{deltapp}
\end{figure}

In fig.~\ref{deltapp}, we present the linewidth of the $s$-$p_x$ intersublevel transition as a function of the energy detuning $\Delta_{pp}$ between the $p$-levels. The linewidth given by the Fermi golden rule for phonon absorption is also shown for comparison. There is a striking difference between the two, which demonstrates the importance of two-phonon processes.
Application of Fermi golden rule results in a overestimate of the linewidth in the vicinity of its maximum efficiency, and a strong underestimate away from this maximum.

Note that we have checked that the assumption of an exponential decay of the ZPL is a good approximation in the full calculation: at 100K, $\text{d}I/\text{d}E < 0.3$. In addition, at the same temperature, the energy displacement $D(E)$ remains smaller than $0.15$~meV.

\begin{figure}
\begin{centering}
\includegraphics[width=0.45\textwidth]{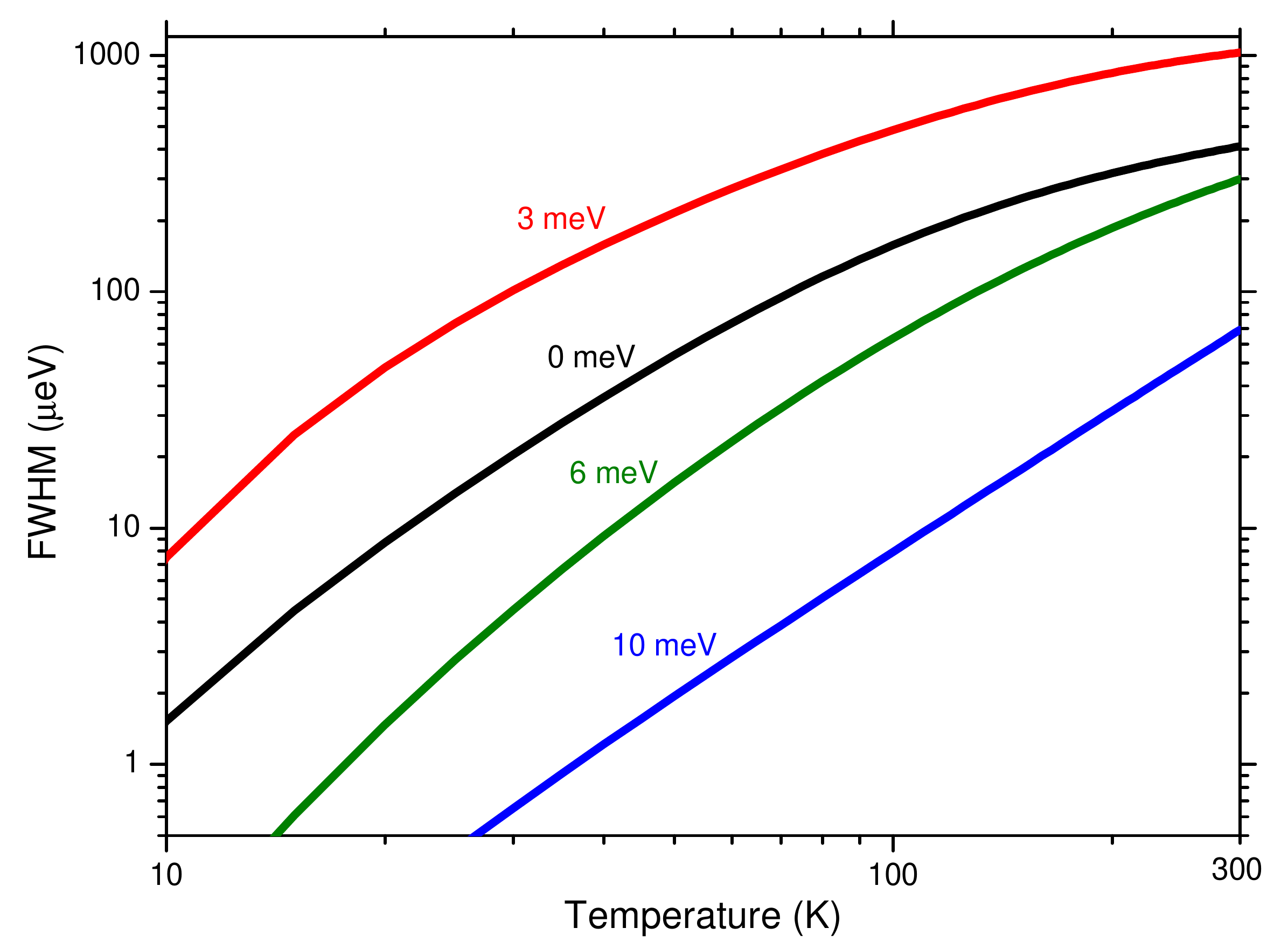}
\par\end{centering}

\caption{(Color online) Linewidth (FWHM) $2\Gamma_0$ of the ZPL of the $s$-$p_x$ intersublevel transition versus temperature for different values of the energy splitting $\Delta_{pp}$ between the $p$-states (the values of $\Delta_{pp}$ are indicated on the figure).}
\label{dephtemp}
\end{figure}

In fig.~\ref{dephtemp}, the $s$-$p_x$ ZPL linewidth as a function of temperature is plotted for different values of the $p_x$-$p_y$ splitting $\Delta_{pp}$. It is worth to mention that the present calculation does not include the population relaxation from $p_x$ to $s$ by anharmonic polaron decay \cite{polaronrelax,sauvage02,zibik04,grange07,zibik09} which contribute additionally to the linewidth, and is the dominant mechanism at low temperature\cite{zibik08}. Note also that the ZPL component in the intersublevel absorption decrease with increasing temperature down to approximatively 0.5 at room temperature.

\begin{figure}
\begin{centering}
\includegraphics[width=0.48\textwidth]{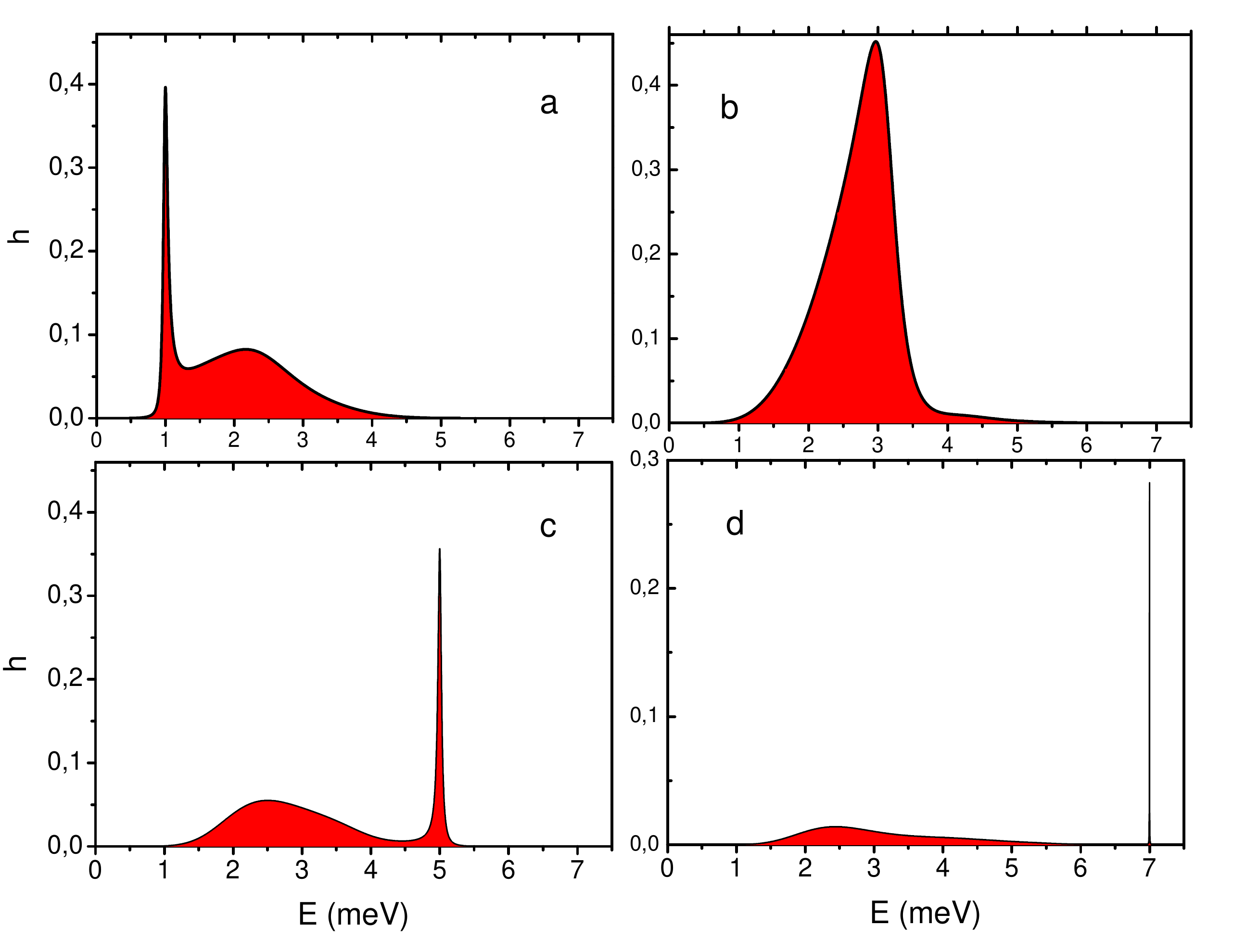}
\par\end{centering}

\caption{(Color online) Function $2h$ representing the contribution to the linewidth $2\Gamma_0^{\alpha}$ as a function of the acoustic phonon energy (T=100K) for different $p$-splittings. (a) 1meV (b) 3 meV (c) 5 meV (d) 7 meV. The area under the curve (red) is equal to $2\Gamma_0$ (FWHM).}
\label{functionh}
\end{figure}

A physical interpretation is provided by studying the integrand (denoted $h$) in Eq.~\ref{galpha} ($\Gamma_0^{\alpha}=\int \text{d} \varepsilon h(\varepsilon)$). $h(\varepsilon) \text{d} \varepsilon$ corresponds to the contribution to the dephasing (non-diagonal terms only) due to phonons with energy comprised between $ \varepsilon - \text{d} \varepsilon/2$ and $ \varepsilon + \text{d} \varepsilon/2$.
In Fig.~\ref{functionh}, the function $h$ is plotted for different value of the $p$-$p$ splitting.
The function $h$ is sharply peaked in $\varepsilon=\Delta$, which corresponds to real transition involving one-phonon process, i.e. phonon absorption promoting the electronic state from $e_0$ to $e_1$. If the condition $\text{d} \gamma_{od}/\text{d}E \ll 1$ is fulfilled (cases c and d), the area of this peak can be approximated by $\gamma_{od}(-\Delta)$ which
corresponds to the Fermi golden rule for phonon absorption. In addition, in the cases a, c and d of Fig.~\ref{functionh}, the function $h$ presents an additional broad peak between 1 and 4 meV which is attributed to 2-phonon virtual processes.

In fig.~\ref{functionh}.b, the contribution to one and two-phonon processes cannot be separated. In this case, the calculated linewidth is found to be smaller to the Fermi golden rule (see Fig.~\ref{deltapp}). This feature can be explained qualitatively by replacing delta energy conservation $\delta(\Delta-\varepsilon)$ appearing in Fermi golden rule by Lorentzian of linewidth $\gamma_{od}(\Delta)$. As a consequence the strong variations in $\gamma_{od}(-\varepsilon)$ are smoothed (over an energy $\gamma_{od}(\varepsilon)$), resulting in a negative correction for the peak of $\gamma_{od}(-\varepsilon)$.

Depending on $\Delta_{pp}$, the relative contribution of real and virtual transitions varies. In Fig.~\ref{functionh}.c, for  $\Delta_{pp}= 5$ meV, the area of the two peaks are of the same order. For larger $p_x$-$p_y$ splitting (Fig.~\ref{functionh}d), dephasing processes are dominated by virtual transitions. This can be explained by noticing that the phonons which are the more efficiently coupled to the electronic states have wavelength of about the dot size, which corresponds to typical energy of 1 or 2 meV in self-assembled QDs. This can be seen by looking at the variation of $\gamma_{od}(-\varepsilon)$ (Fermi golden rule for absorption) on fig.~2. In case phonons with QD wavelength are not resonant with the electronic transition energy, they cannot induce real transitions, but can still be elastically scattered by the QD and consequently induce decoherence. On the contrary, the efficiency of real absorption processes decreases rapidly when increasing $\Delta_{pp}$. This qualitatively explains why virtual transitions can be the dominant dephasing mechanism.

\section*{Conclusion} 

We have presented a non-pertubative calculation of acoustic phonon induced dephasing of optical transitions in QDs. The Hilbert space has been limited to states differing by up to 2-phonons from the initial state, and the calculation has been made non-pertubatively within this framework.
An analytic formula has been given in terms of the electron-phonon matrix elements. The theory has been illustrated to the study of the dephasing of intersublevel transitions in QDs. The contribution of phonons to the dephasing has been studied as a function of their energy.

\section*{Acknowledgment}
The author is very grateful to R. Ferreira and G. Bastard for many useful discussions.
The author acknowledges support from the Alexander von Humboldt foundation.

\appendix
\section{Acoustic phonon sidebands}

In order to calculate the polarization decay, we decompose the initial phonon state $|\phi_0\rangle$ on the new phonon eigenstate $|\widetilde{\phi_i}\rangle $:
\begin{equation}
\begin{split}
\langle e_0,\phi_0 | e^{-iHt/\hbar}|e_0,\phi_0\rangle = \sum_{\widetilde{\phi_i}} |\langle \phi_0|\widetilde{\phi_i}\rangle|^2 \langle e_0,\widetilde{\phi_i} | e^{-iHt/\hbar}|e_0,\widetilde{\phi_i}\rangle \\
 + \sum_{\widetilde{\phi_i}}\sum_{\widetilde{\phi_j}\neq\widetilde{\phi_i}} \langle \phi_0|\widetilde{\phi_i}\rangle \langle \phi_j|\widetilde{\phi_0}\rangle \langle e_0,\widetilde{\phi_i} | e^{-iHt/\hbar}|e_0,\widetilde{\phi_j}\rangle
\end{split}
\end{equation}
The second term is of the order of $\mathcal{O}(1/N_{cr})$ and thus vanishes.
The states $|\widetilde{\phi_i}\rangle$ involved in the above equation differs from the  state $|\widetilde{\phi_0}\rangle$ (corresponding to the zero-phonon line) by a finite number of phonon occupancies, so that in the limit $N_{cr} \rightarrow \infty$ we have:
\begin{equation}
\langle e_0,\widetilde{\phi_i} | e^{-iHt/\hbar}|e_0,\widetilde{\phi_i}\rangle = e^{-i(\widetilde{\varepsilon_i}-\widetilde{\varepsilon_0})t/\hbar}\langle e_0,\widetilde{\phi_0} | e^{-iHt/\hbar}|e_0,\widetilde{\phi_0}\rangle
\end{equation}
This allows us to write:
\begin{equation}
\langle e_0,\phi_0 | e^{-iHt/\hbar}|e_0,\phi_0\rangle =  P_Z(t)g(t)
\end{equation}
\begin{equation}
P_Z(t)= \langle e_0,\widetilde{\phi_0} | e^{-iHt/\hbar}|e_0,\widetilde{\phi_0}\rangle
\end{equation}
\begin{equation}
g(t)=  \sum_{\widetilde{\phi_i}} |\langle \phi_0|\widetilde{\phi_i}\rangle|^2 e^{-i(\widetilde{\varepsilon_i}-\widetilde{\varepsilon_0})t/\hbar}
\end{equation}

The function $g(t)$ defined here is the Fourier transform of 
\begin{equation}
g(E)=  \sum_{\widetilde{\phi_i}} |\langle \phi_0|\widetilde{\phi_i}\rangle|^2 \delta(E-(\widetilde{\varepsilon_i}-\widetilde{\varepsilon_0})) ,
\end{equation}
which is the function describing the acoustic phonon sidebands absorption as a function of the energy detuning with the zero-phonon line.

As result from the independent boson model theory \cite{mahan,vagov02,vagov03}, $g(t)$ can be expressed as:
\begin{subequations}
\begin{equation}
g(t)=\exp\left[\int_{-\infty}^{+\infty}\text{d}E f(E) \left( e^{-iEt/\hbar}-1\right) \right] 
\end{equation}

\begin{equation}
\begin{split}
f(E) = \sum_{\phi_i} \frac{|M_{i}^{e_0e_0}-M_{i}^{gg}|^2}{E^2} \delta(|E| - \varepsilon_i)
\\ \times \begin{cases} (N(E)+1) & \text{if $E>0$} \\ N(-E) & \text{if $E<0$} \end{cases}
\end{split}
\end{equation}
\end{subequations}
The function $g(t)$ describes the evolution of the coherence due to phonon sidebands only, \textit{i.e.} taking into account diagonal electron-phonon interaction only. The acoustic phonon sidebands have been studied extensively in the literature for interband transitions in both energy domain \cite{besombes01,favero03} and time domain \cite{vagov04}, and more recently for intraband transitions\cite{zibik08}.


\end{document}